\documentclass[12pt,a4paper]{article}
\usepackage[T2A]{fontenc}
\usepackage[cp1251]{inputenc}
\usepackage[russian,english]{babel}
\usepackage{amssymb}
\usepackage{amsmath}
\usepackage{graphicx}
\usepackage[small,sc,center]{titlesec}
\usepackage{indentfirst}
\usepackage[labelsep=period]{caption}
\usepackage{caption}
\newcommand{\msun}{\,$M_{\odot}$}
\newcommand{\msyr}{\,$M_{\odot}$\,yr$^{-1}$}
\newcommand{\rsun}{\,$R_{\odot}$}

\newcommand{\ergs}{\,erg\,s$^{-1}$}
\newcommand{\gcmq}{\,g\,cm$^{-3}$}

\newcommand{\gcmsq}{\,g\,cm$^{-2}$}
\newcommand{\kms}{\,km\,s$^{-1}$}

\newcommand{\cmsqg}{\,cm$^2$\,g$^{-1}$}

\newcommand{\ha}{H$\alpha$}

\begin{document}
	
\begin{center}
\textbf{\large \ha\ clarifies picture of IIn supernova SN~2015da}
	
	\vskip 5mm
\copyright\quad
2026 \quad N. N. Chugai \footnote{email: nchugai@inasan.ru}\\
\textit{Institute of astronomy, Russian Academy of sciences, Moscow} \\
Received  30.05.2026
\end{center}

{\em keywords:\/} stars -- supernovae; supernovae -- SN~2015da; supernovae -- dust formation

{\em PACS codes:\/} 
\clearpage
 
 \begin{abstract} 

Observational data on type IIn supernova SN~2015da are revisited to clarify some ambiguous  
   issues of this phenomenon. 
The \ha-emitting site is found to switch around  day 90 
  from the circumstellar matter to the unshocked ejecta.
This picture makes it possible to recover the expansion velocity of late time ejecta
 and to locate the dust formation in the inner zone of unshocked ejecta. 
The bolometric light curve and expansion velocities combined with modelling  the  circumstellar interaction 
  imply that the supernova exploded  with an energy of $\gtrsim 4\times10^{51}$ erg  inside the circumstellar envelope with the mass of $\approx 14$\msun.
The recovered density of the circumstellar matter is consistent with the Thomson optical depth 
 inferred from \ha\ modelling on day 46.
The observed circumstellar envelope is produced by an average mass loss rate of 0.035\msyr\ during 
  400\,yr before the supernova explosion.
 	
\end{abstract} 	

\section{Introduction}

Supernovae SN~IIn is a category of diverse phenomena unified by the presence of a dense circumstellar (CS)
  matter (CSM); "n" stands for the narrow emission lines in their spectra (Schlegel 1990).
These are rare events and thus are rather distant.
The initial stage after the explosion, therefore, is usually missed that hampers the understading of the  overall physical picture. 

The supernova IIn  SN~2015da is an exception from this rule.
It was discovered soon after the explosion and was followed spectroscopically and photometrically for 8 years 
 (Tartaglia et al. 2020: Smith et al. 2024, hereafter T20 and  S24, respectively).
Remarkably, the spectral evolution has been traced in detail at the early stage, which turns out below a key 
 factor in clarifying a physical picture of phenomena similar to SN~2015da.   

The interpretation of observational data on SN 2015da  (T20; S24) suggests that this SN~IIn is characterized by a large mass of the CSM estimated from $5.5-10.5$\msun\ (T20) to 20\msun\ (S24), and a high explosion energy of $(5-10)\times10^{51}$ erg implied by a large radiated energy     
 (S24).
Inferred mass and a time of CS shell formation suggest tremendous  average mass loss rate up to 
  $\sim 0.1$\msyr\ (S24) that could be provided by either LBV pulsations (T20) or by a mass loss from a common envelope of a massive binary (S24).

The large volume of observational data, especially at the early stage,
 motivates us to revisit the SN~2015da issue in attempt to  clarify some uncertain points, particularly, the CSM mass, explosion energy, and the location the source of the dust infrared (IR) flux emerged after about day 500 (T20). 
 
Three options for the origin of the IR dust emission are conceivable: 
 (i) IR echo from the pre-existing 
   dust (T20; Fransson et al. 2014) similar to SN~1979C and SN~1980K (Dwek 1983); (ii) dust formation in the cold dense shell (CDS) between the forward and reverse shock (S24) following 
    conjecture for SN~1998S (Pozzo et al. 2004); (iii) dust formation in the undisturbed SN envelope  likewise in SN~2010jl (Maeda et al. 2013).
       
In the present paper, in Section \ref{sec:general}, the \ha\ observations are analysed to 
  conclude that at late times the broad \ha\ emission originates in the unshocked ejecta.
 In Section \ref{sec:model} the CS interaction model is used to infer  
 the CSM mass and the SN explosion energy based on the light curve and the SN  expansion velocity.
The recovered CSM density is supported by the model of the CS \ha\ emission (Section \ref{sec:thomson}).
The Section \ref{sec:dust} provides an answer to the question on the new dust location.

\begin{figure}
	\centering
	\includegraphics[trim=0 100 0 130, width=\columnwidth]{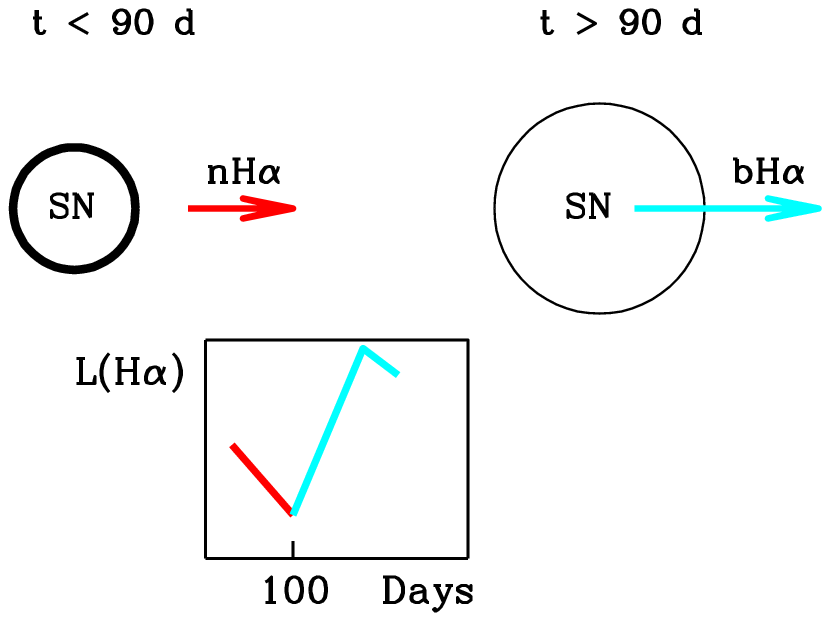}
	\caption{\footnotesize
		Visualisation of the two-stage evolution of SN~2015da with 
	 the luminosity of CS \ha\ ({\em red}) and \ha\ from SN ejecta ({\em cyan }) 
	 schematically shown in the square. 
	 At the early stage  the observed narrow \ha\ (n\ha) with the decreasing luminosity forms in the CSM;
	 at this stage the emission from SN envelope is fully blocked by the 
	 opaque CDS.
	 At the late stage  the CDS becomes transparent and we see the broad \ha\ emission with the increasing luminosity originated from the unshocked SN ejecta, whereas the CS \ha\ emission is already weak.	
		}
	\label{fig:cart}
\end{figure}
%

\section{Interpretation of \ha\ evolution}
\label{sec:general}

One can identify two physically different stages of the SN~2015da spectral evolution   (T20). 
At the early stage $t < t_{tr} \approx 90$ days the spectrum of SN~2015da is a smooth continuum with a prominent \ha\ emission (narrow core and broad wings).
 The narrow component is emitted by the CSM, whereas broad wings are due to  
   a Thomson scattering of line photons in the same CSM with the Thomson optical depth 
  $\tau_{\tiny {\rm T}} \gtrsim 1$.  
The opaque CDS fully blocks the radiation from the unshocked SN ejecra,
  so we do not see broad lines from the undisturbed SN, the phenomenon formerly recognized in SN~1998S (Chugai 2001).
The termination of this stage is manifested by the weakening CS \ha\ emission and transition to the late stage $t > t_{tr}$  that is characterized by the emergence of broad \ha\ emission 
with growing luminosity emitted by the unshocked SN ejecta.
 The transition is presumably related to the 
 significant drop of the CDS hydrogen opacity caused by the decreasing effective temperature. 

The transition from the early to the late stage is well illustrated by a sharp V-like fall/rise of 
 the \ha\ luminosity accompanied by the rapid increase of the line width 
   (T20, Fig. 14, Table 5) from FWHM = 1450\kms\ on day 79 to 
    FWHM = 3770\kms\ on day 138.
    The \ha\ maximum luminosity  of $1.4\times10^{42}$\ergs\ is attained on day 210 (T20).

The identification of the broad \ha\ emission with unshocked ejecta at $t > t_{tr}$ has two important implications. 
  Firstly, it permits us to measure the maximal velocity of the 
 unshocked ejecta from the blue edge of \ha\ emission.
Particularly, on day 138 the \ha\ (T20) shows the maximal ejecta velocity 
  of $\approx 6500$\kms, 
  while on day 2603 (S24) the expansion velocity is of $\approx 3000$\kms. 
Secondly, it is of high significance for the issue of the new dust location (Section \ref{sec:dust}).

Noteworthy, the presented picture of the dominant role of the unshocked SN ejecta 
  in the late \ha\ luminosity is an alternative option with respect to the 
  \ha\ emitted by the CS clouds in the forward shock proposed for 
   SN~2010jl (Chugai 2018). 
  This model is also plausible, since it is based on the formation of a broad 
  velocity spectrum of \ha-emitting gas as result of the crushing and fragments acceleration  of CS clouds in the forward shock (Klein et al. 1994).
However, the absence of a broad \ha\ shock wave component at the early stage 
 is a strong argument against such origin of the \ha\ emission at the late stage.

\begin{table}
	\vspace{6mm}
	\centering
	{{\bf Table 1.} CSM mass and model parameters}\\
	\bigskip	 
	\begin{tabular}{p{3cm}p{2cm}p{2cm} p{2cm} } 
		\hline	
		Параметр                    & m5     &  m10  &  m20        \\
		\hline	
		
		$M_{ej}$  [\msun]             &  5  &  10    &  20           \\
    	$ E $     [$10^{51}$\,erg]    & 4       & 5    &  7.3         \\	
	  $M_{cs}^{\bigstar}$  [ \msun]  &  14.8  &  13.9      &  12.7      \\
			$M_{ni}$  [ \msun]           & 0.07   &  0.07       &  0.07      \\
		$r_1$  [$10^{16}$\,cm]      &  1.2    &    1      &    1 \\
		$r_2$ [$10^{16}$\,cm]      &  2.8       & 2.5     &  2.7    \\               
		$k_1$                     &  -0.6    &  -0.6      &  -0.6      \\ 
		$k_2$                     &  -2.5    &  -2.5      &  -2.6              \\
		$k_3$                     &  -2     &  -2        &  -2                \\
		\hline  
		 	\parbox[]{6cm}{$^{\bigstar}$\small{Mass inside the radius $10^{17}$\,cm.}}	     
		\end{tabular}
\end{table}
%

%
\section{Mass of CS envelope}
\label{sec:model}

The CS interaction is modelled based on the thin shell hydrodynamics (Chevalier 1982).
Kinetic luminosity of each shock (forward and reverse) is converted into X-ray emission 
 with the conversion factor $\eta = \mbox{min}(1, t/t_c)$, where $t_c$ is the cooling time 
 calculated using the cooling function (Sutherland and Dopita 1993).
The X-ray power of each shock wave absorbed by the cold gas (unshocked ejecta, CDS, CSM) in total constitute the model bolometric luminosity.
The absorbed X-ray power is calculated using an absorption coefficient for the solar composition $\kappa = 84(E/1\,\mbox{кэВ})^{-8/3}$\cmsqg\ (Longair 2011).

Supernova with the kinematics $r = v/t$ is adopted to have the density distribution 
at the initial moment $\rho = \rho_0/[(x(1 + x^7)]$, where $x = v/v_0$, while 
 $\rho_0$ and $v_0$ are determined by the ejecta kinetic energy $E$ and mass $M_{ej}$.
The adopted SN density distribution qualitatively reflects a general features of  SN~1987A ejecta:
 a relatively flat distribution in the inner zone $\rho \propto v^{-1}$ 
 (Imshennnik and Nadezhin 1989;  Chevalier and Soker 1989) 
    and a steep drop $\rho \propto v^{-8}$ in outer layers; our principal results are not sensitive to the external power index. 

The CSM density is set by a broken power law  $\rho \propto r^k$ with the inflection radii $r_1$ and $r_2$ 
and power index $k_1$ for $r < r_1$, $k_2$ for $r_1 < r < r_2$, and $k_3$ for $r > r_2$.
The model CSM is distributed in the range berween the preSN radius $R_0 = 150$\rsun\ and   
  the model shock radius at $t = 3030$ days ($10^{17}$\,cm), the epoch of a last observation of SN~2015da (S24).  

\begin{figure}
	\centering
	\includegraphics[trim=0 70 0 150, width=\columnwidth]{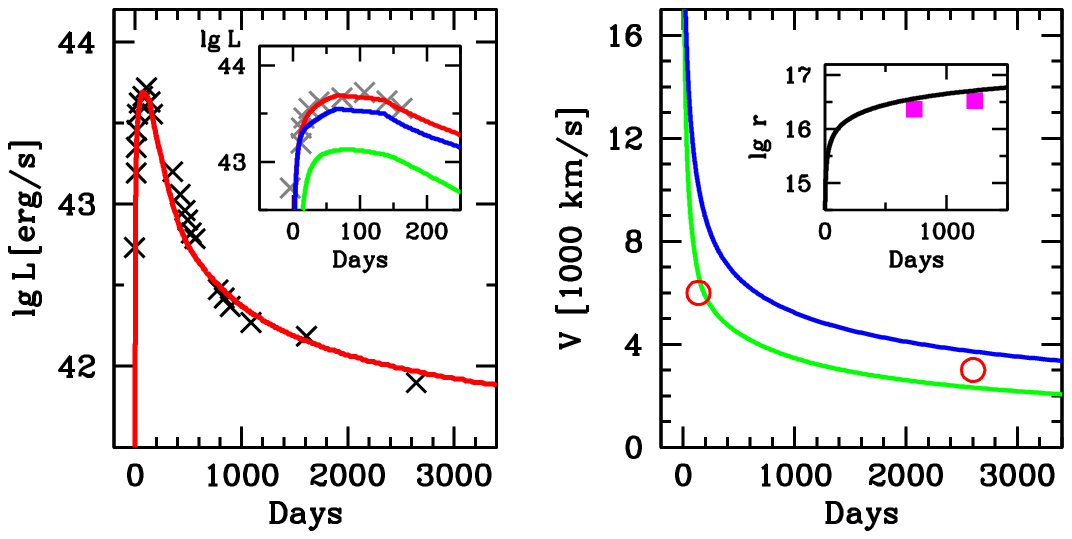}
	\caption{\footnotesize
		{\em Left.} Bolometric light curve ({\em red}) for the model with the ejecta mass of 5\msun\ (Table 1) 
		overplotted on the observational light curve ({\em crosses}) from (S24).
		Inseret shows the region of the maximum together with contribution of forward shock 
		({\em blue} line)	 and reverse shock ({\em cyan}).	 
		{\em Right.} Maximum velocity of the unshocked ejecta ({\em blue} line) and  the CDS velocity 
		({\em cyan}) along with the observational velocity of blue edge of \ha\ emission.
		Insert shows the CDS radius ({\em line}) along with the black body radius of the dust 
		({\em squares}) on days 733 and 1233 (T20).
	}
	\label{fig:int3}
\end{figure}
%

\begin{figure}
	\centering
	\includegraphics[trim=0 70 0 150, width=\columnwidth]{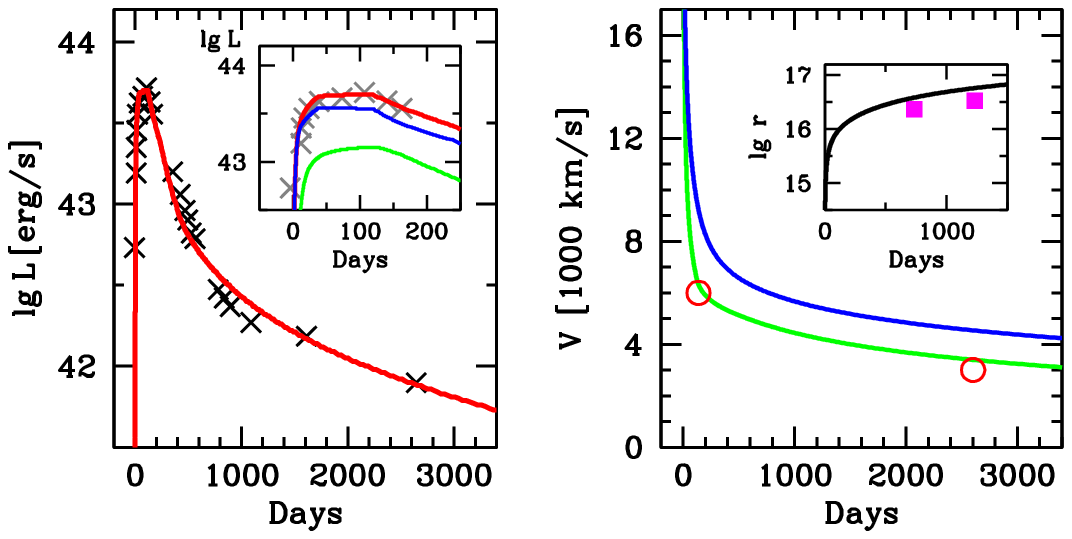}
	\caption{\footnotesize
		The same as Figure  \ref{fig:int3}, but for the 20\msun\ ejecta.
	}
	\label{fig:int2}
\end{figure}
%
Three models presented in Table 1  have ejecta mass of 5, 10, and 20\msun.
The explosion energy is a free parameter that is inferred by the optimal fit to the light curve (S24) and SN expansion velocity. 
The parameters of the CSM turn out practically the same for all the models, with relatively minor scatter around mean values. 
Shown in Figures \ref{fig:int3} and \ref{fig:int2} are models m5 and m20; the plot for m10 model is omitted because it is very much similar to presented models.
All the models reproduce the bolometric light curve (S24) and ejecta maximal velocities on days 138 and 2603 
 to a satisfactory degree.
A major contribution in the luminosity is provided by the forward shock (Fig. \ref{fig:int3} and \ref{fig:int2}, left insert).
 This explains, why for all the models the CSM mass is rather similar. 
The models include  0.07\msun\ of $^{56}$Ni that actually does not affect noticeably the overall luminosity. 

Despite of the decrease of the explosion energy with decreasing ejecta mass, 
  models with the ejecta mass  significantly smaller than  5\msun\ are not able to fit observations 
  with the  energy $< 4\times10^{51}$ erg; the latter value 
  should be considered, therefore, as a minimum value of the SN~2015da
  explosion energy. 
Noteworthy, this value significantly exceeds a possibility
of the neutrino-driven explosion ($< 2\times10^{51}$ erg (Janka 2017)).
The remaining possibilities  are either magnetorotational explosion 
 (Bisnovatyi-Kogan 1971), or collapsar-driven explosion 
 (rapid disk accreation,  onto a black hole) (Woosley 1993).

The right insert (Fig. \ref{fig:int3} and \ref{fig:int2}) shows CDS radius evolution compared to 
   observatonal estimates of the dust black body radius  with the temperature of 1400\,K and 1010\,K on 
    day 733 and 1233, respectively (T20).
Noteworthy, for both moments the dust black body radius is smaller than the CDS radius by the same factor of 
 $\approx 1.6$.
For the maximal ejecta velocity at this stage of about 5000\kms\  (Fig. \ref{fig:int3} and \ref{fig:int2}), therefore, the IR emission could originate from the dust formed in the ejecta inner zone with velocities $\lesssim 3000$\kms.
This conjecture is supported below by the \ha\ profile modelling (Section \ref{sec:dust}).

The presented CS interaction model suggests the instant radiation escape, although at first glance tremendous 
  CSM mass might result in the large radiation diffusion time comparable to the time of the bolometric light 
    maximum $\sim 50-100$ d (T20, S24). 
In reality, at the maximum $t_{max} \approx 50$\,d the  CSM column density in the 
 CS interaction model is 3\gcmsq\ that is translated to  
   the Thomson optical depth of $\tau_{\tiny \rm T} \approx 1$ assuming a complete ionization.
  The diffusion time $t_{dif} \approx \tau r/c$ thus turns out to be comparable to the light travel time $r/c\approx 2$ days, 
    substantially less than $t_{max}$, which justifies the instant escape approximation.

\section{\ha\ and CSM density}
\label{sec:thomson}

The density distribution of the CSM inferred via the CS interaction 
  model can be tested using the \ha\ profile at the early stage, when this line forms in the CSM.
To this end we model the \ha\ line profile in the Keck/DEIMOS spectrum on day 46 (T20).
The spectrum has been used earlier to demonstrate that narrow \ha\ with broad wings forms in the 
  CS shell with the large Thomson optical depth $\tau_{\tiny {\rm T}} \gg 1$ (T20). 
 
Here we apply a model that apart from the Thomson scattering takes into account 
 the resonance scattering in \ha. 
The resonance scattering brings about two effects. Firstly, the resonance scattering of the CDS continuum 
   radiation produces an additional P Cyg component. Secondly, the resonance scattering can force line photons to experience larger amount 
    of Thomson scatterings.
  As result, a moderate Thomson optical depth $\tau_{\rm T} \sim 1$ turns out sufficient       
    to describe \ha\ broad wings.    
  
The computational zone of CSM provided by the CS interaction model lies between the optically thick CDS at  $r_s = 5.2\times10^{15}$\,cm on day 46 and the external radius of $9r_s$. 
The electron number density is a fraction ($x$) of the model CS hydrogen density 
 ($n_{\rm \tiny H}$), i.e., $n_e = xn_{\rm \tiny H}$. 
 The adopted electron temperature is $10^4$\,K that is close to the temperature of the hydrogen collisional ionization.
The recombination \ha\ emissivity is $j =An_e^2$ ($A$ is a normalizing factor).
 The \ha\ Sobolev optical depth along a direction $\vec{s}$ is $\tau(r,\mu) = Q|du_s/ds|^{-1}$, where $Q$ is a free parameter assumed to be independent of the radius, whereas the velocity gradient depends on the CSM kinematics $u(r)$ and cosine $\mu$ of an angle between the photon wave vector and radius. 
  The radiation transfer is computed by the Monte Carlo technique.
 
Modelling shows that the assumption of the constant CSM velocity 
 $u_w =  90$\kms\ (S24) does not permit to describe the profile at the transition between the narrow core and 
  broad wings; the CSM velocity close to the CDS of a radius $r_s$ should be somewhat larger.
 This requirement is implemented via the superposition of the constant velocity and the pre-acceleration component  $u(r) = u_{ps}\exp[-b(r/r_s-1)] + u_w$, where 
  $u_{ps} = 170$\kms, $b = 4$, and  $u_w =  80$\kms.
The emissivity close to the CDS $r_s < r < 1.2r_s$ should be  also
 larger compared to approximation $j = An_e^2$ by a of factor of 2.5.
 
\begin{figure}
	\centering
	\includegraphics[trim=0 70 0 150, width=\columnwidth]{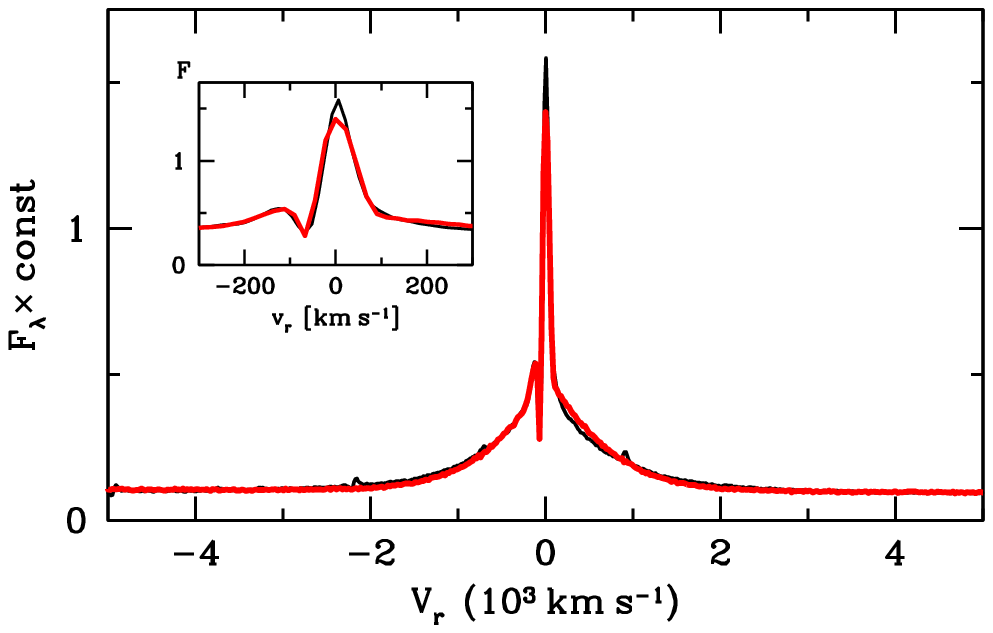}
	\caption{\footnotesize
		\ha\ in the spectrum of SN~2015da on day 46 ({\em black} line) along with 
		the optimal model ({\em red} line).
		Insert shows fragment in the velocity range $\pm300$\kms; the effect of high velocity of the CSM and 
		larger emissivity close to CDS is seen as a flux excess in the radial velocity range -100... -200\kms.
	}				
	\label{fig:ha}
\end{figure}

The optimal \ha\ description on day 46 is found for $\tau_{\tiny {\rm T}} = 1.1$  (Fig. \ref{fig:ha}),
which corresponds to the hydrogen ionization fraction $x = 0.9$ for the density of the CS interaction model.  
This can be considered as a good agreement between the CSM density inferred from tha CS interaction model and the CS density implied by the \ha\ Thomson wings.

So far we ignored that the model  \ha\ luminosity on day 46 of $4.4\times 10^{40}$\ergs\ is significantly  
 (9 times) lower than the observational value of $4\times 10^{41}$\ergs\ (T20).
This dramatic disparity is likely due to a clumpy structure of the CSM. with 
 a clumpiness parameter 
 $C = \langle\rho^2\rangle/\langle\rho\rangle^2 \approx 9$ .
  This raises a question, whether the approximation of smooth density distribution  
  is valid for the modelling the \ha\ profile; the issue is addressed in the Appendix.

\section{Dust in supernova envelope}
\label{sec:dust}

In Section \ref{sec:model},  when comparing black body radius of the dust (T20) with the CDS radius,
   	I suggested that the source of the dust emission is the inner zone of SN ejecta expanding at about 3000\kms.
This assumption is checked here via the modelling blueshifted \ha\ on day 810 
  (T20).	  

The model assumes the envelope with the kinematics $v = r/t$ and 
  the boundary velocity of 4500\kms. 
The \ha\ emissivity distribution is  $j = A\exp{(-q(v/v_b))}$, where $A$ is a   
  normalizing factor, $q$  determines the line width and its value 
    is adjusted from the blue part of the \ha\ profile. 
 The dust is distributed homogeneously on average in the sphere  $v < v_d$.
 Yet the dust resides in the form of opaque clouds of a radius $a$  and a small filling factor $f\ll 1$, but with a large occultation optical depth  $\tau_{oc} = (3/4)fv_dt/a > 1$, somewhat similar to 
   SN~1987A (Lucy et al. 1991).
The \ha-emitting hydrogen in the dusty zone occupies the intercloud space.   

\begin{figure}
	\centering
	\includegraphics[trim=0 100 0 100, width=\columnwidth]{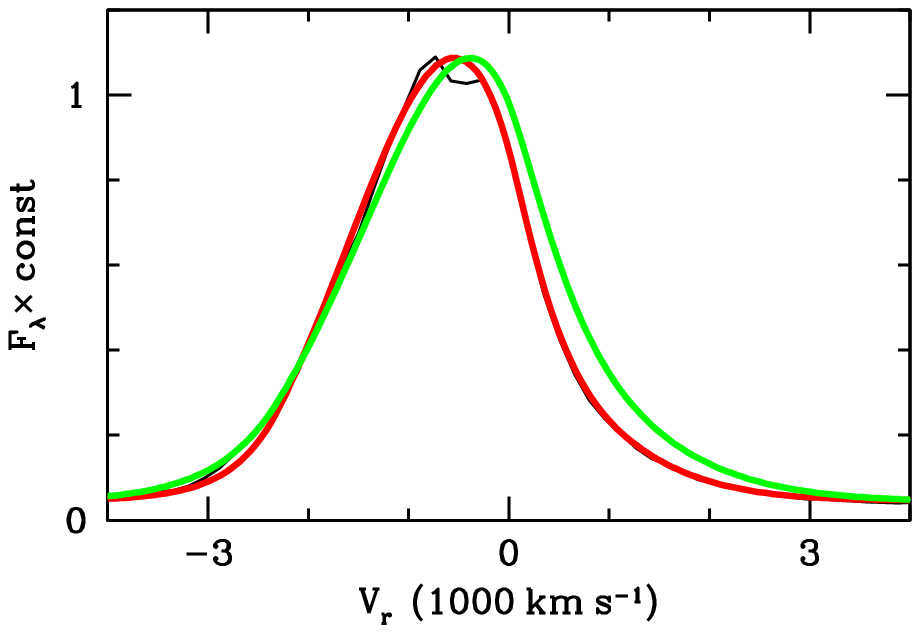}
	\caption{\footnotesize
	 \ha\ on day 810 ({\em black } line) (T20) with overplotted optimal model of dusty ejecta with the dust  inside the sphere expansing at 2300\kms\ and optical depth of 3.3 ({\em red}). 
	 The {\em green} line shows the case of the dust optical depth of 2.	
	 					 	}
	\label{fig:dust}
\end{figure}

The best fit model (Fig. \ref{fig:dust}) describes the blueshifted \ha\ for the parameter values: 
$v_d = 2300\pm200$\kms, $q = 12.3\pm0.4$, and $\tau_{oc} = 3.3\pm0.2$.
The case of the dust optical depth $\tau_{oc} = 2$ and $q = 9.5$ is shown to demonstrate the sensitivity to the dust optical depth.
The \ha\ escaping luminosity of the best-fit model is attenuated by 7 times
  which is consistent with the drop \ha\ luminosity by a factor of 7 between days 500 and 750 
  (T20, Figure 14), when presumably the dust forms.  
 The performed modelling thus supports the picture of the new dust formation in the inner zone of unshocked ejecta.

Remarkably, the absence of strong red wing of \ha\ at this stage indicates the low effective albedo 
 that is a feature of an ensemble of opaque dusty clouds. 
The dust in unshocked ejecta is an alternative with respect to the model that suggests the dust formation 
  only in the CDS. 
It should be emphasised that in the picture of \ha\ emitted by the unshocked ejecta the absorption 
   in the dusty CDS cannot produce the \ha\ blueshift at all.   
  
The low limit of the dust mass that is needed to provide the optical depth of $\tau_d = 3$ is
 $M_d =  (16/9)\pi R^2b\zeta\tau_d \sim 2\times10^{-5}$\msun, where the adopted radius of the dusty zone on day 810 
 is  $R = v_dt = 1.7\times10^{16}$\,cm, the radius of the dust grain $b = 10^{-5}$\,cm, and 
  the density of dust grain material $\zeta = 1$\gcmq.

%
%

\section{Discussion}
\label{sec:discus}

The aim of the paper has been to estimate CSM mass and explosion energy    
    of type IIn SN~2015da and clarify some uncertain points of the physical picture.  
   
I find that the evolution of the \ha\ luminosity and the line width presented by Tartaglia et al. (2020) 
  strongly suggests that at $t = t_{tr} \approx 90$ day after the explosion
 the \ha-emitting site switches from the CSM to the unshocked ejecta. 
This behaviour is consistent with the blocking SN ejecta emission by the opaque  
  CDS at the early stage ($t < t_{tr}$) and subsequent clearing the CDS at the later  
   stage ($t > t_{tr}$).
This picture was recognized in SN~1998S (Chugai 2001)
  and described in detail for SN~2023ixf (Utrobin and Chugai 2026).
 The similar transformation of \ha\ width, from narrow to broad, is shown by the 
 spectral evolution of type IIn  SN~2024by (Goto et al. 2026).  
 
The conclusion on the origin of the late time \ha\ emission in the unshocked ejecta 
  is of importance in two respects. 
First,  in this picture the blue edge velocity of \ha\ emission is a measure of 
the maximum velocity of the unshocked SN ejecta; this velocity is used 
in the model of the CS interaction  to recover the mass of the CSM. 
Second, this picture permit us to use late time \ha\ for testing the conjecture 
on the dust formation in the unshocked SN ejecta at $t \gtrsim 500$ days.

 The association of the late time \ha\ emission with the unshocked  SN ejecta is 
   an alternative to the model of SN~2010jl (Chugai 2018), where 
   the  dominant \ha\ emission was attributed to CS clouds crushed, fragmented 
     and accelerated in the forward shock (Klein et al. 1994).
 
The CS interaction model combined with the observational light curve and expansion velocity  permits us to determine the CSM mass of  $\approx14$\msun\ and a low limit of the explosion energy $\gtrsim 4\times10^{51}$\,erg.
Remarkably that the found CSM mass varies in the range of only 7\% 
  around the mean value for explored range of SN ejecta 5 - 20\msun.
The inferred density of the CSM is supported by the modelling of the CS \ha\ on day 46.
 
The CS \ha\ modelling reveals an effect of the CSM acceleration ahead of the CDS 
 up to $\approx 170$\kms\ that is beyond the acceleration by the SN radiation ($\approx 40$\kms). 
  The additional $u_{ps} = 130$\kms\ could be due to the CSM acceleration by a cosmic ray (CR) pressure. 
 The required CR acceleration efficiency (i.e., the ratio of CR pressure at the shock wave to the 
 dynamical pressure) is $\eta = p_c/\rho v_{cds}^2 
 \approx 2u_{ps}/v_{cds}$ (Chugai 2019), which for $u_{ps} = 130$\kms, and the CS interaction model value $v_{cds} = 9200$\kms\ on day 46 results in  $\eta \approx 0.03$.
       
The expansion velocity of the CSM $u = 80$\kms\  and the outer CSM radius $R = 10^{17}$\,cm indicate that 14\msun\ has been lost during $t_{loss} = R/u \approx 400$\,yr before the explosion with the average  mass loss rate of  $\approx 3.5\times10^{-2}$\msyr.
The enormous mass loss rate of pre-SN~2015da could take place in the common envelope (CE)  regime
  of a massive binary (S24; Chugai and Danziger 1994).

A conceivable CE scenario for 10\msun\ SN ejecta suggests the preSN of about 
 26\msun\ before losing of 14\msun.
  During 400 years before the core collapse preSN was at the carbon burning phase
 that spans $\sim 10^3$\,yr with the RSG radius of $R_1 \approx 1400$\rsun\ (Woosley et al. 2002).
Since the luminosity and radius of the RSG steadily increase with time it is conceivable that about 400 years before the core collapse a secondary component of massive binary system with the major semiaxis 
  $a \sim R_1$ could find itself in the CE of the RSG with a subsequent loss of large mass fraction.

Until now we did not touch a problem of a high \ha\ luminosity on dau 210.
To illustrate the point, let us consider a homogeneous sphere of ionized hydrogen 
 with the radius $r = vt = 1.1\times10^{16}$\,cm (where $v = 6000$\kms) 
   and volume $V$.
  For the electron temperature of 6000\,K the \ha\ effective recombination coefficient 
  $\alpha_{32} = 1.9\times10^{-13}$\,cm$^3$\,s$^{-1}$ in Case B (Osterbrock and Ferland 2006).
  The \ha\ luminosity $L = 1.4\times10^{42}$\ergs\  (T20) implies then 
   the Thomson optical depth $\tau_{\tiny \rm T} = 
    \sigma_{\tiny \rm T}r(L/V\alpha_{32}h\nu)^{0.5} = 4.9$/
That large  Thomson optical depth unavoidably would result in the large \ha\ blueshift ($\sim 3000$\kms) inconsistent with the observed \ha\ at this stage (T20). 
 The solution of the \ha\ luminosity problem might be a clumpy structure of the SN ejecta that could 
  substantially decrease the effective Thomson optical depth, but retain the high \ha\ luminosity.
This picture, however, deserves a separate study elsewhere.

\section{Conclusion}

We conclude  with a summary of major results.
\begin{itemize}
	
	\item {The \ha\ evolution before and after day 90 reflects the transition from the stage of 
		 \ha\ emission by the CSM towards the \ha\ emission by the unshocked SN ejecta.}
		\item { The bolometric light curve and SN expasion velocity imply the shock interaction 
			between the SN ejecta with the CS envelope of $\approx 14$\msun.}
		\item  { The SN explosion energy is $\gtrsim 4\times 10^{51}$\,erg, beyond the neutrino-driven  mechanism.}
		\item {The observed CSM is produced by the mass loss at the average rate of 0.035\msyr\ during 
			the last 400 years before the SN explosion.}
		\item {The \ha\ emission on day 46 forms in the CSM with the Thomson optical depth of $\approx 1.1$ 
			that is consistent with the density distribution recovered by the CS interaction model.}
	   \item {The \ha\ blueshift after the day 500 is an outcome of the \ha\ absorption by the dust
	   	formed in the inner zone of the SN ejecta.}
\end{itemize}

\bigskip   
\section{Acknowledgements}

I thank Leonardo Tartaglia for sending me SN~2025da spectra.


\bigskip
\begin{center}
{\large Appendix}

\medskip
 {\large Homogeneous approximation for clumpy envelope}

\end{center}

\bigskip

Let the envelope with the local average density $\rho$ 
consists of clouds with the filling factor $f$, mass fraction $\eta$, and an intercloud medium.
The cloud radius $a$ is adopted to be significantly smaller compared to the scale of average density variation $r_s$.
The density of cloud and intercloud matter is 
\begin{equation}
	\rho_c = \frac{\eta}{f}\rho\,\qquad \rho_{ic} = \frac{1 - \eta}{1 - f}\rho\,.
	\label{eq:density}	
\end{equation}
Emission measure of this two-component medium is larger than that of homogeneous medium 
by a clumpiness factor 
\begin{equation}
	C = \eta^2/f + (1 - \eta)^2/(1 - f)\,.
\end{equation}
Assuming that masses of cloudy and intercloud components are equal, i.e., $\eta=0.5$, 
  one gets the clumpiness factor to be $C = [4f(1-f)]^{-1}$.
In that case the condition $C = 9$ implies cloud filling factor $f \approx 0.03$.

The optical depth of the clumpy medium at the length $l$ is the sum of cloudy and intercloud components  
\begin{equation}
	\tau = \frac{fl}{s}[1 - \exp(-\tau_c)] + (1 - f)r\kappa \rho_{ic}\,,
\end{equation}
where $s$ is an average secant of a cloud, $\tau_c =\kappa \rho_c s$ is the average optical depth of a cloud, $\kappa$ is the   coefficient of the Thomson scattering.
In the approximation of spherical clouds one has $s = 4a/3$ and $\tau_c = (4/3)\kappa \rho_ca$.
The first term in the right-hand side of  Eq. (3)  is a product of the average number of clouds on the length 
$l \sim r_s$ (i.e. the occultation optical depth (Lucy et al. 1991)) and the the probability of the scattering in the cloud.  
Here we use an approximation    
 $\langle 1 - \exp{(-\tau)}\rangle \approx 1 - \exp{(-\langle\tau\rangle)}$.
In the limit of optically thin clouds $\tau_c \ll 1$ taking into account expressions (\ref{eq:density}) one finds the optical depth on the length $l = r_s$ to be  $\tau = \kappa\rho r_s$, the same 
 as in the homogeneous case with the average density.
Angular variations of optical depth will be small, if along any directions of photon propagation 
  the number of clouds large, i.e., the occultation optical depth  $\tau_{oc} = (3/4)fr_s/a > 1$.
This suggests an upper bound of the cloud radius $a/r_s < f = 0.03$. 

The Thomson optical depth of clumpy medium with the characteristic scale of $r_s$, therefore, is the same as  
 for the homogeneous medium with the average density, if clouds are optically thin and the radius of clouds is small, $a < fr_s$.

\clearpage

\section{References}

\noindent
Bisnovatyi-Kogan G. S., Soviet Astron.  {\bf 14},  65 (1971)\\
\medskip
Chevalier R. A., N. Soker N.,  Astrophys. J. {\bf 341}, 867 (1989)\\
\medskip 
Chevalier R. A., Astrophys. J. {\bf 259}, 302 (1982)\\
\medskip
Chugai N. N., Danziger I. J.,  Mon. Not. R. Astron. Soc. {\bf 268}, 173 (1994)\\
\medskip
Chugai N. N., Mon. Not. R. Astron. Soc. {\bf 326}, 1448 (2001)\\
\medskip
Chugai N. N., Mon. Not. R. Astron. Soc. {\bf 481}, 3643 (2018)\\
\medskip
Chugai N. N., Astron. Lett.{\bf 45}, 71 (2019)\\
\medskip
Dwek E., Astrophys. J. {\bf 274}, 175 (1983)\\
\medskip
Fransson C., Ergon M., Challis P. J. et al., Astrophys. J. {\bf 797}, 118 (2014)\\
\medskip
Goto S. Yamanaka M. Kawabata K. S., arXiv260618585G (2026)\\
\medskip
Imshennik I. S., Nadezhin D. K., Sov. Sci. Rev., Sect. E, {\bf 8}, Part 1, p. 1 (1989)\\
\medskip
Janka H.-T., {\em Handbook of Supernovae}. Springer International Publishing AG, p. 1095 (2017)\\
\medskip
Klein R. I., McKee C. F., Colella P., Astrophys. J. {\bf 420}, 213 (1994)\\
\medskip
Longair M. {\em High Energy Astrophysics.} Cambridge, UK: Cambridge University Press (2011)\\
\medskip
Lucy L. B., Danziger I. J., Gouiffes C., Bouchet P.
{\em Supernovae. The Tenth Santa Cruz Workshop in Astronomy and Astrophysics.}
Ed. S.E. Woosley. New York: Springer-Verlag, 1991\\
\medskip
Maeda K., Nozawa T., Sahu D. K.,  Astrophys. J. {\bf 776}, 5 (2013)\\
\medskip
Osterbrock D. E, Ferland G. J.,  {\em Astrophysics of gaseous nebulae 
	and active galactic nuclei}. Sausalito, CA: University Science Books (2006) \\
\medskip
Pozzo M., Meikle W. P. S.  Fassia A. et al., Mon. Not. R. Astron. Soc. {\bf 352}, 547 (2004)\\
\medskip
Schlegel E., Mon. Not. R. Astron. Soc. {\bf 244}, 269 (1990)\\
\medskip
Smith N, Andrews J., Milne P. et al., Mon. Not. R. Astron. Soc. {\bf 530}, 495 (2024)\\
\medskip
Sutherland R. S., Dopita M. A., Astrophys. J. Suppl. {\bf 88}, 353 (1993)\\
\medskip
Tartaglia L., Pastorello A.,  Sollerman J. et al., Astron. Astrophys. {\bf 635}, 
id.A39, 22 pp (2020)\\
\medskip
Utrobin V. P., Chugai N. N., arXiv260615395U (2026)\\
\medskip
Woosley S. E., Heger A., Weaver  T. A., Review Mod. Phys. {\bf 74}, 1015 (2002)\\
\medskip
Woosley S. E., Astrophys. J. {\bf 405}, 273 (1993)\\

\end{document}